# Polarization-controlled non-Hermitian metasurfaces for ultra-sensitive terahertz sensing

Xintong Shi [1], Hai Lin [1, *], Tingting Liu [2,3], Yun Shen [4, *], Rongxin Tang [4], Le Li [4], Junyi Zhang [1, 4], Yanjie Wu [1, 5], Shouxin Duan [4], Chenhui Zhao [4], Shuyuan Xiao [2, 3, *]

1* College of Physics Science and Technology, Central China Normal University, Wuhan 430079, China

2 School of Information Engineering, Nanchang University, Nanchang 330031, China

3 Institute for Advanced Study, Nanchang University, Nanchang 330031, China

4 Institute of Space Science and Technology, Nanchang University, Nanchang 330031, China

5 School of Electronic Science and Engineering, University of Electronic Science and Technology of China, Chengdu, 611731 China

*Corresponding author(s). E-mail(s): linhai@mail.ccnu.edu.cn; shenyun@ncu.edu.cn; syxiao@ncu.edu.cn

## Abstract

Non-Hermitian systems offer significant advantages in sensor design, especially at the exceptional points. However, the extreme sensitivity near these points poses great challenges due to fabrication errors and system noises, which degrade sensing performance. To address this, we introduce a novel approach leveraging the polarization degrees of freedom in non-Hermitian systems. In this work, we establish a direct relation between the incident polarization and the transmission phase of a coupled metasurface system and achieve the polarization-controlled phase singularity even post-fabrication. The incident polarization angle can be utilized as a sensing index, which enables indirect and accurate measurement. The theoretical approach is experimentally validated using a general design of THz non-Hermitian metasurface sensors. Our method enhances robustness and sensitivity, opening new avenues for practical applications in ultra-sensitive sensing.

## Introduction

Any natural, non-isolated physical systems are intrinsically non-Hermitian, because they continually exchange energy with the surrounding environment[1,2]. The non-Hermitian phenomena have been observed across various fields such as optics, acoustics, microwaves, and electronics [3–9]. The intriguing characteristics of these phenomena have inspired numerous exciting applications[10–13], for example, non-Hermitian state-based topological lasers [14], unclonable physical keys [15], and more flexible wavefront manipulation and topological engineering capabilities [16,17]. Additionally, leveraging the eigenvalue splitting near exceptional points (EPs） [18–20], which are among the most renowned exotic states in non-Hermitian systems, can facilitate the development of high-performance sensors with diverse photonic structures like optical resonators, microcavity, and metasurfaces [21–29]. Among them, metasurfaces are highly favored due to their unprecedently flexible and versatile design strategies [30–37], and in particular provide feasible solutions in previously inaccessible THz regime. Their significantly enhanced local field [38–40], combined with the unique characteristics of THz waves, present tremendous advantages in detecting subtle biological signals [41–43].

Despite EPs' superior performance, a critical issue has been consistently overlooked in the existing design strategies for previous non-Hermitian sensors. The ultra-sensitivity near the EPs, offering both benefits and drawbacks. Eigenvalue variations under a perturbation are not a good measure of the overall sensor performance near an EP [44]. The factors causing eigenvalue splitting include not only the perturbation from the analytes but also fabrication errors, which would lead to a significant degradation in sensing performance. Considering the geometric parameters are fixed once the structures are fabricated, some studies have introduced gain media to compensate for uncertainty [45–47]. However, such approach is not suitable for periodic structures like metasurfaces, let alone the lack of such gain media in the terahertz regime. This issue appears to be a deadlock, severely obstructing the practical application of non-Hermitian metasurface sensors.

In this work, we resort to the polarization degree of freedom to achieve controllable singularity in non-Hermitian systems for sensing. We theoretically construct an effective Hamiltonian[48] by incorporating the coupling efficiency loss which comes from the mismatch between the linearly polarized light and the excited eigenmodes of the structure. By solving its eigenvalues, we derive the explicit relation between the polarization of incident light and the transmission phase of coupled system, and observe the phase singularity in the polarization space. This phase singularity exhibits high tolerance to fabrication errors. Based on this, the polarization of incident light, which can be easily tuned for phase singularity even after fabrication, is utilized as a sensing index. Finally, the theoretical method is experimentally demonstrated using a general design of THz metasurface sensors and measured with the THz time domain spectrometer. The proposed method explores the new degree of freedom in non-Hermitian physics, and open an avenue for robust ultra-sensitive sensing.

# Results

## Theoretical derivations

In previous studies, non-Hermitian metasurfaces have typically relied on circular polarization or strictly linear polarization. Experimental measurements of the linear polarization or synthesized circular polarization usually require strict horizontal and vertical polarization to evaluate co-polarization and cross-polarization. Here we introduce a sensing method that incorporates polarization degrees of freedom.

Figure 1 shows the schematic illustration of the proposed non-Hermitian metasurface sensor and its working mechanism. The unit cell of the metasurface sensor is constructed by two coupled and orthogonal resonators. The resonators possess similar frequencies but different intrinsic rates, respectively. A linearly polarized light is incident upon the metasurface with a controllable polarization angle $\theta$, and a detector collects signals corresponding to the horizonal and vertical polarizations aligned with two resonators, respectively. By rotating the first polarizer to control the incident polarization angle, the excitation strength of two resonators can be precisely tuned. Due to the complementary relation between the incident polarization angle and the orthogonal resonators, the transmissions in horizontal and vertical directions exhibit a distinct behavior. One exhibits phase anomaly and the other does not. For the anormal one, we also observe that the coupled system is expected to exhibit a polarization-controlled phase singularity of transmission (which is extremely sensitive to any subtle change in the surrounding environment) at a certain angle. Consequently, the sensing strategy here, constructs the method to characterize the analyte using mechanical control for incident polarization. The subfigures illustrate a phenomenon where the polarity of group delay reverses, providing a clearer representation of phase singularity. On either side of the critical polarization angle, the calculation of the winding number for the transmission spectrum reveals that the transmission signals corresponding to the incident polarizations exhibit different topological invariants. This discrepancy serves as the origin of the phase singularity. Next, we proceed to construct the equivalent Hamiltonian.

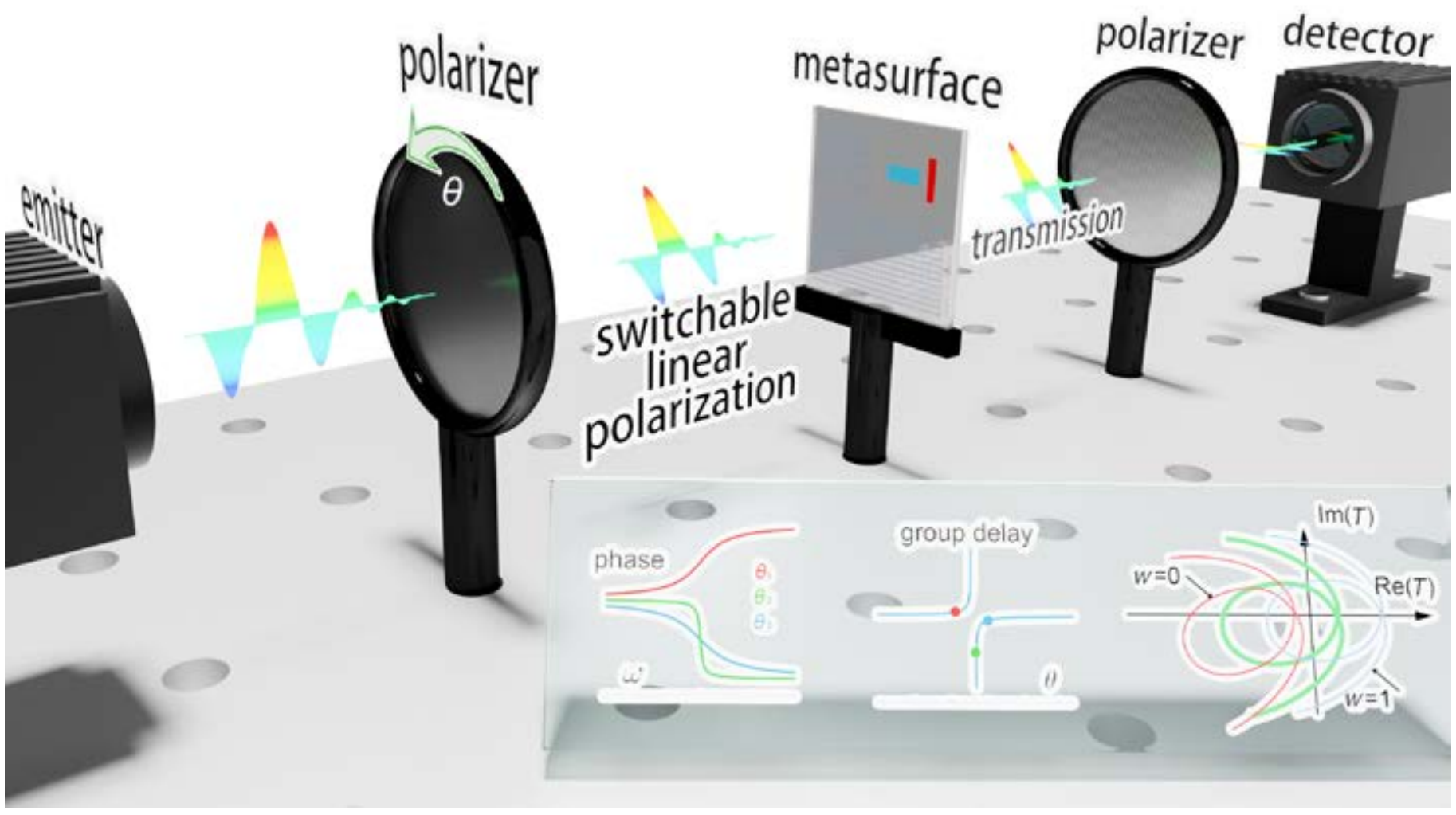

Fig 1. The conceptual illustration of the proposed non-Hermitian metasurface sensor and working mechanism. The unit cell consists of two orthogonal resonators. A pair of polarizers, functioning as a polarizer and an analyzer respectively, controls the polarization of the incident light and analyzes the polarization of transmission into two directions along the resonators (horizontal and vertical directions). The phase changes from lagging to leading, the corresponding group delay exhibits a transition, and the topological invariant of transmission also changes.

For an isolated resonator, when the direction of incident light aligns with that of its eigenmode, it experiences the strongest excitation and harvests the highest energy, and when the incident polarization directions are orthogonal, it acquires the lowest energy. In a more general case, if the incident polarization directions form an angle, it would lead to losses between the resonator and the incident field. For a coupled system composed of two resonators, the situation becomes more complex as the resonators can exchange energy with each other. If the two resonators are orthogonal, the losses that come from their coupling with the incident light would change in opposite ways. We bring these polarization-induced losses into consideration (for more information please see supplementary Note SN1). The effective Hamiltonian of the coupled system can be written as

$$H = \begin{bmatrix} \omega_1 - i\gamma_1 - i\Gamma_1 & \kappa e^{i\beta} \\ \kappa e^{i\beta} & \omega_2 - i\gamma_2 - i\Gamma_2 \end{bmatrix}, \tag{1}$$

where $\omega_{1,2}$ are the frequency of the resonators, $\gamma_{1,2}$ are the intrinsic loss of the resonators, and $\Gamma_{1,2}$ are the polarization-induced losses. The coupling strength in coupled system is denoted as $\kappa e^{i\beta}$. The complex frequencies are denoted $\Omega_1$ and $\Omega_2$ , and the corresponding eigenvalues can be denoted as

$$\lambda_{\mp} = \bar{\Omega} \pm \sqrt{\Delta\Omega^2/4 + \kappa^2 e^{2i\beta}}, \tag{2}$$

where $\bar{\Omega} = \frac{\Omega_1+\Omega_2}{2}$, and $\Delta\Omega = \Omega_1 - \Omega_2$. It is obvious that the two eigenvalues coalescing at EP with $\lambda_{\mp} = \bar{\Omega}$ when the equation satisfying with the discriminant of a quadratic equation $\Delta\Omega^2/4 + \kappa^2 e^{2i\beta} = 0$.

While the physics of EPs may appear straightforward, it is often challenging to reach the point in practice because EPs require quite stringent structural parameters. Therefore, in the following, we introduce tuneable losses by leveraging the polarization degree of freedom, which allows us to dynamically control the coupled system response.

Therefore, the eigenvalues are re-written as

$$\lambda_{\mp} = \overline{\Omega} \pm \sqrt{k^2 cos2\beta + \Delta\omega^2/4 - (\Delta\gamma + \Delta\Gamma)^2/4 - i\kappa^2 sin2\beta - i\Delta\omega(\Delta\gamma + \Delta\Gamma)}\,, \quad (3)$$

where $\Delta\Gamma$ is the difference between polarization-induced losses of the resonators. Considering the trend of this difference, the system responses should have an extremum (for more information please see supplementary note SN1 and Figure S3). It is obvious to find the zero occurs at $\overline{\Omega} - \sqrt{\Delta\Omega^2/4 + \kappa^2 e^{2i\beta}}$ and $\overline{\Omega} + \sqrt{\Delta\Omega^2/4 + \kappa^2 e^{2i\beta}}$.

Assuming that the coupled system satisfying the conditions of zero, $\overline{\Omega} = \sqrt{\Delta\Omega^2/4 + \kappa^2 e^{2i\beta}}$. Under such condition, the Magnitude $M_{\lambda_} = 0$, and the phase expression is,

$$\varphi_{\lambda_} = \arg(\lambda_) = \arctan\frac{-\omega_1(\Gamma_2+\gamma_1) - \omega_2(\Gamma_1+\gamma_2) - \kappa^2 sin2\beta}{\omega_1\omega_2 - (\gamma_1+\Gamma_1)(\gamma_2+\Gamma_2) - \kappa^2 cos2\beta}. \quad (4)$$

The phase expression has a singularity point when the denominator resembles a critical coupling condition ($1 - \alpha^2 - \kappa^2 = 0$). If the two resonators share the same eigenfrequency, the total losses equal the energy exchanged between the two resonators. Both of the real and imaginary parts $\lambda_$ are zero, which means the non-transmission appears along the direction of this resonator.

Figure 2a illustrates the parameter space of the coupled system under tight coupling ($\beta = 0$) as functions of frequency difference $\Delta\omega$ and incident polarization angle $\theta$, as indicated by the equation 2. The upper and lower panels show the real and imaginary parts of eigenvalues, respectively. The real eigenvalues not only separate as the frequency difference increases, but are also influenced by the incident polarization, with the effect on the real eigenvalues diminishing as the frequency offset between the two resonators increases. The ring appears on the $\theta$ axis when the two resonators share the same eigenfrequency ($\Delta\omega = 0$).

The imaginary part of eigenvalue is mainly controlled by the incident polarization. When the polarization angle is distant from the quadrant center (45° ), the imaginary eigenvalues separate. Otherwise, the imaginary eigenvalues degenerate. When the frequency offset increases, the imaginary eigenvalues exhibit only an intersection point. Otherwise, they degenerate into a line along the $\theta$ axis.

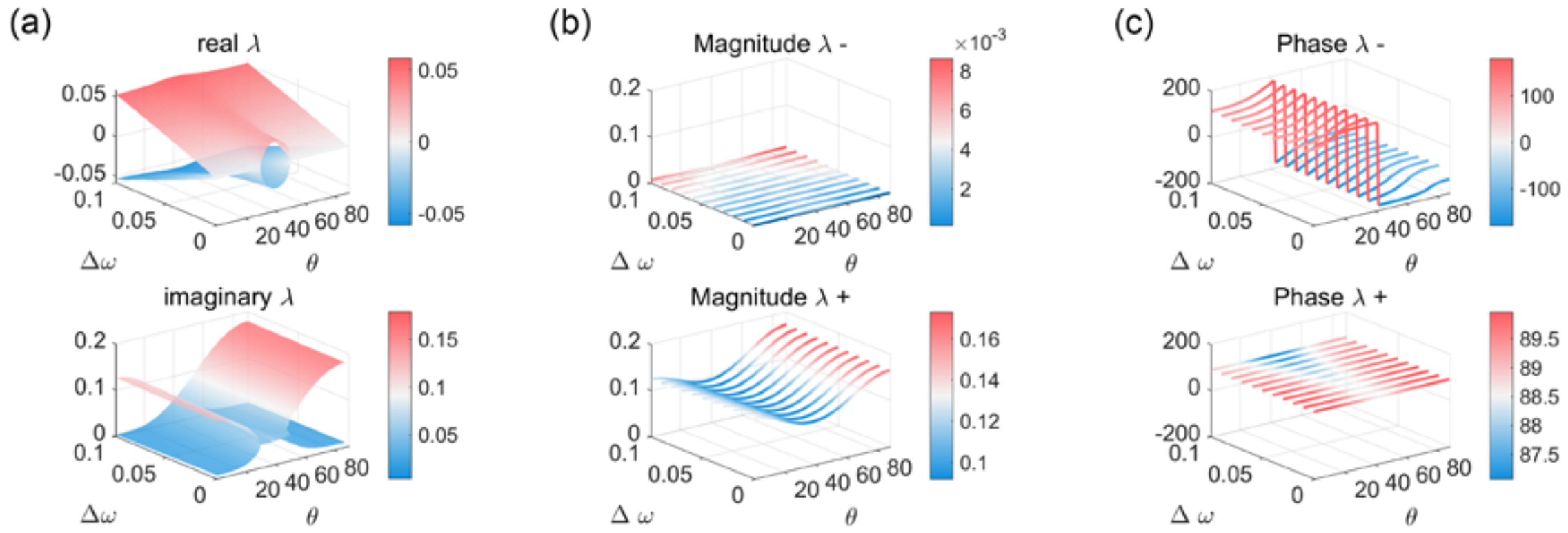

Fig. 2 The theoretical calculations (a). The parameter space of coupled system with phase difference $\Delta\omega$ and incident polarization angle $\theta$. The upper panel and down panel are real part and imaginary part of eigenvalues of the effective Hamiltonian. The two parts of eigenvalues form a closed loop and a line close to critical polarization angle, respectively. (b-c). The Magnitudes of two eigenvalues and the phases of two eigenvalues. The results are calculated by the equations (2) and (4). The parameters used for the calculations are $\gamma_1 = 0.2$, $\gamma_2 = 0.13$, and $\kappa e^{i\beta} = 0.03e^{i0}$.

In this work, we concentrate on the system's phase singularity. Figures 2b and 2c show the calculated magnitude and phase of eigenvalues based on equation 4. The magnitude of $\lambda_-$ goes to the minimum, approaching to zero (see the legend of 2b), and the phase exhibits a singularity, when $\Delta\omega$ equals zero. The phase changes from lagging to leading when the incident polarization crosses the critical angle. The magnitude of $\lambda_+$ is much larger than $\lambda_-$, and the phase is trivial.

It is worth noting that the magnitude and phase in Figures 2b and 2c correspond to the eigenvalues of the effective Hamiltonian, representing the effects of the Hamiltonian operator on the incident field. These values do not include the propagating phase and therefore represent the intrinsic (non-spectral) phase of the effective Hamiltonian rather than the directly measured phase in the spectra. In addition, even if there is a certain offset between the eigenfrequencies of two resonators, the phase singularity can still exist, albeit appearing at different incident polarization angles. It is important to note that the phase singularity is influenced by the incident polarization, thus establishing a polarization-controlled singularity system.

Consequently, this relation allows us to establish a sensing function with respect to the incident polarization angle. The optical response of the coupled system would change after a small amount of analyte is added onto the metasurface. In equation 4, the resonance frequencies $\omega_{1,2}$ (main) and intrinsic losses$\gamma_{1,2}$ (very small in the THz regime) are influenced by the perturbations, but the polarization-induced losses $\Gamma_{1,2}$ depend on the incident field. Then the equation $\omega_1{}'\omega_2{}' - \kappa^2 cos2\beta = (\gamma_1 + \Gamma_1)(\gamma_2 + \Gamma_2)$ determines the new incident angle. Hence, it bridges the two aspects of polarization and analyte, allowing us to characterize the perturbation in the polarization space, which links the electromagnetic parameters of the metasurface with external mechanical parameter.

## Design and simulation results

We employ the finite integration technique (FIT) to validate the proposed theoretical method. The metasurface sensor consists of simple and classical structures, namely split ring resonators (SRRs) and cut wires (CWs). The selection of these simple structures ensures manageable responses while maintaining sufficient coupling effects. The parameters of the structures, including the SRRs and CWs, are estimated using the ruler equation [49]. By combining these two structures into a single unit, a tightly coupled system is constructed. Figure 3a provides a schematic illustration of the metasurface configuration. We conduct simulations to obtain the phase response of the metasurface without any analytes under different incident polarizations. The designed parameters of the metasurface sensor are presented in Table 1(for more information please see supplement information SN2 and SN3), and the corresponding results are plotted in Figure 3b. The upper panel shows the transmission phase of horizontal direction, and the lower panel shows the transmission phase of vertical direction. The responses have an approximate frequency offset of 0.11 THz. The responses are consistent with the theoretical calculations: one exhibits a phase singularity, while the other is trivial.

Table 1 Parameters of the non-Hermitian THz metasurface sensor

| Parameters | Symbols | Values (μm) |
|---|---|---|
| Length of SRR in the *x*-direction | $L_{1x}$ | 60 |
| Length of SRR in the *y*-direction | $L_{1y}$ | 50 |
| Length of CWs | $L_2$ | 70 |
| Width of SRR | $W_1$ | 10 |
| Width of CWs | $W_2$ | 10 |
| Gap of SRR | $G$ | 10 |
| Distance between SRR and CWs in the *x*-direction | $d_x$ | 42.5 |
| Displacement of SRR from the center of unit cell in the *y* direction | $d_{\mathrm{y}}$ | 20 |

We proceed with the phase singularity. It can be observed that under different linearly polarized excitation, the transmitted phase lags on one side of the singularity point, and upon crossing the singularity, on the other side, the lagging phase transitions to leading. This behaviour is consistent with the results shown in Figure 2. What distinguishes this from the calculations in Figure 2 is that here we have the real phase of the resonators, including the propagation phase. The phase singularity appears at 42° and 0.827 THz. The phase variation becomes pronounced near the singularity point. We also investigate the topological origin used winding number for this phase singularity, as included in supplementary information SN5.

Retrieving the phase singularity becomes challenging when considering the influence of propagation phase, and it is also difficult to characterize it as a single point. From an engineering perspective, analyzing the differential form of the phase is much more convenient. When the extremum of the group delay ($d\varphi/d\omega$) undergoes a transition, it indicates that the phase has crossed the singularity. We also illustrate the corresponding group delays of the two directions in the inset of Figure 3b. In other words, the extremum of the group delay corresponds well to the phase singularity. Moreover, since the derivative of the propagation phase with respect to frequency is constant, its effect is directly eliminated in this process.

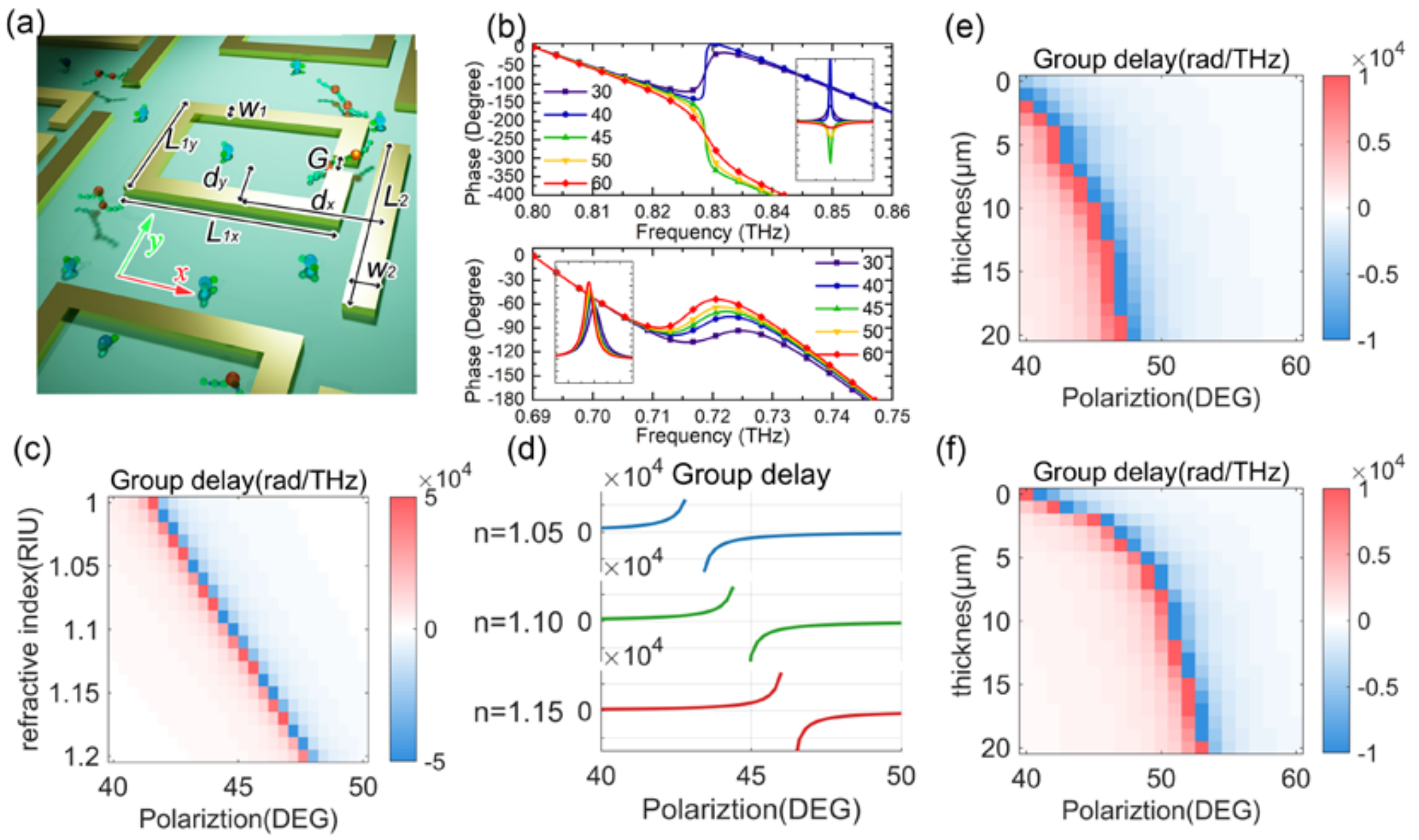

Fig. 3 Simulation results. (a). The schematic of the non-Hermitian THz metasurface sensor. (b). The transmission phase (group delay) of the metasurface without loading analyte, (c). The relation between the incident polarization angle (*x*-axis) and refractive index (*y*-axis), with group delay indicated by color. (d). The three refractive indexes are extracted from Figure 3c to illustrate the results. (e) and (f) are the group delay with varying the thickness of analytes, with the refractive index are 1.2 and 1.4, respectively.

In the FIT simulations, a 20 μm dielectric layer is introduced onto the metasurface sensor to analog the analyte [50]. The range of analyte refractive index considered is from 1 to 1.2, while the incident polarization angle is varied between 40 and 50 degrees. The simulation results are presented in Figure 3c, where each pixel corresponds to once simulated group delay (for more information please see supplement information SN4). The horizontal axis represents the incident polarization angle, while the vertical axis represents the refractive index of the dielectric layer. The color scale indicates the extreme values of the group delay. Notably, a distinct transition is observed, indicating a relation between the two variables characterized by a series of singularities. Beside the transition region, the group delay exhibits different symbols, with one side being positive and the other side being negative. This transition signifies that within the given range, each value of the refractive index corresponds to a unique incident polarization angle. Thus, the result demonstrates the existence of a specific correlation between the refractive index and the incident polarization angle within a defined range.

To illustrate this relation clearly, three specific refractive index parameters (1.05, 1.1, and 1.15) are selected from Figure 3c. Figure 3d depicts the corresponding group delay values for these refractive indexes. The behavior of the three lines aligns with the theoretical predictions. Notably, at the incident polarization angles of 43°, 45°, and 46°, the group delays undergo a transition from positive to negative, revealing the presence of phase singularities. In essence, these findings demonstrate that the incident polarization angles can effectively represent the refractive index.

However, it is important to note that the refractive index of analytes in different concentrations is not the only factor affecting terahertz biosensing. The effective refractive index varies depending on the properties of the substrate and the attached material. Additionally, terahertz detection often involves the removal of moisture, further complicating the use of refractive index alone for accurate terahertz biosensing. To address this, it is necessary to simulate different concentrations of analytes by employing dielectric blocks with varying thickness.

Here, we simulate the metasurfaces coated with a dielectric layer of constant refractive index, with the thickness ranging from 0 to 20 μm. Two types of dielectric layers with distinct refractive indexes are used to emulate different analytes. The corresponding results are presented in Figure 3e and Figure 3f, revealing that the sensitivity to the thickness parameter is non-linear. While the thickness variation of the dielectric layer is linear, the corresponding incident polarization angle exhibits a non-linear response. Moreover, as the dielectric layer becomes thicker, the incident polarization angle undergoes slower changes. Additionally, the incident polarization angle varies differently depending on the type of dielectric layer, demonstrating the sensor's capability to specifically distinguish between different analytes. Furthermore, the system's response is more sensitive when a thinner dielectric layer is employed, and the curves corresponding to different refractive indexes (representing different analytes) exhibit distinct patterns. Based on this principle, we attribute the sensor with the capability to accurately identify different analytes.

## Experimental validations

Finally, we also conduct experimental demonstrations of the metasurface sensors to validate the proposed method. The analyte used in the experiments is aflatoxin B1 with varying concentrations. During the measurement process, a switchable linearly polarized terahertz wave is illuminated onto the metasurface sensor at a specific polarization angle $\theta$. The incident polarization angle is precisely controlled by a rotating polarizer. The transmitted signal is then separated into the horizontal axis using a polarizer and subsequently detected by a receiver.

Additionally, we implement the key measure to enhance measurement precision. To minimize the need for frequent adjustments of the polarizer angle, multiple metasurfaces are fabricated, each coated and dried with aflatoxin B1 at different concentrations. This process plays a crucial role in improving the accuracy of measurement. The calibration procedure involved setting the first polarizer to have the same polarization as the photoconductor antenna, while adjusting the polarization of the second polarizer to be orthogonal to the first. The second polarizer is carefully aligned to achieve calibration when the transmission amplitude reaches its minimum. The angle of the first polarizer is then set by selecting points at 5-degree intervals around 45 degrees. After each setting, the unload sensor and other sensors with varying aflatoxin B1 concentrations are sequentially placed on the light path, and a set of responses is measured.

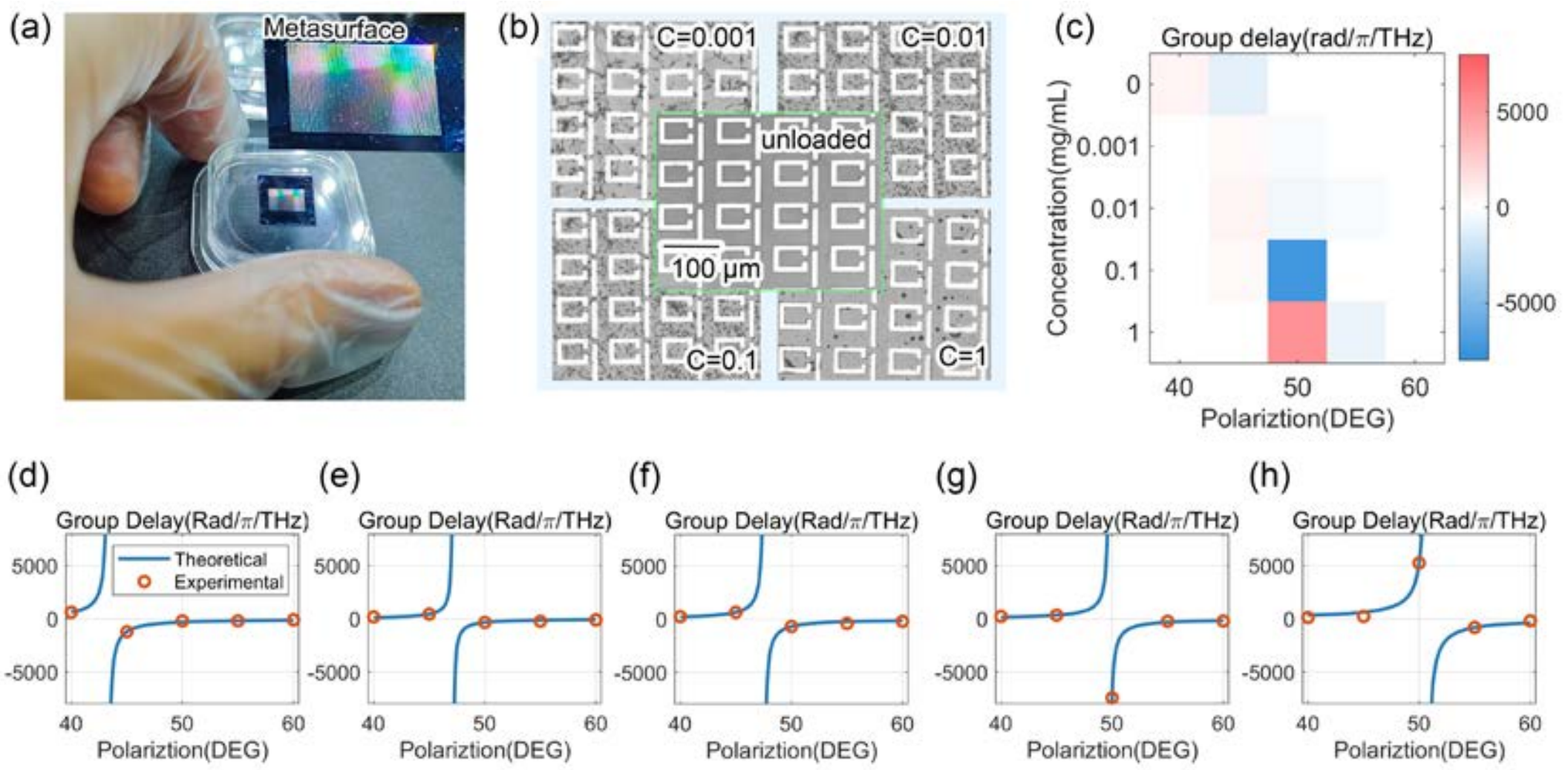

Fig. 4. Experiment results. (a). The fabricated sample of metasurface sensor. (b). The magnified imaging of metasurface sensors coated with aflatoxin B1 of different concentrations, (c). The group delay calculated using a set of 5×5 measurements, revealing the relation between incident polarization angle and concentration of analytes. (d-h). The fitted group delay based on measurement data.

Figure 4a presents a photograph of the fabricated sensor. Figure 4b illustrates magnified photographs of the metasurface sensors coated with different concentrations of aflatoxin B1 (unloaded, 0.001, 0.01, 0.1, and 1 μg/mL, respectively). In these images, the gray region represents the silicon substrate, the white region corresponds to the aluminum structure, and the black dots represent the analytes. The processed group delay as shown in Figure 4c (the raw data pleases see supplementary information SN 6), the same method depicts the group delay between incident polarization and concentration of aflatoxin B1. The results have verified the relation, it is as similar as the anticipation of theoretical and simulation results.

The phase transition of the unload sensor is observed between 40 and 45 degrees. For sensors with concentrations of about 0.001μg/mL, 0.01μg/mL, and 0.1μg/mL, the phase transitions appear between 45 and 50 degrees. Sensors with a concentration of 1μg/mL of aflatoxin B1 exhibit a transition between 50 and 55 degrees. Based on the proposed theory, we can solve the singularity point of each concentration.

In fact, it is not necessary to precisely measure each of group delay under different incident polarization angles. The location of the phase singularity can be determined through multiple measurements on an experimental platform that has not been specially modified. The mathematical form of the phase related to the incident polarization has been established in equation 4. The constant parameter can be determined by switching the incident polarization and measuring the corresponding responses without loading any analyte. Subsequently, when a perturbation is introduced into the system, the new polarization angle associated with the singularity can be determined. Hence, the location of phase singularity can be fitted based on the measured data, when the analytes are added onto the metasurface. Then, the entire group delay with the incident polarization from 40 to 60 degrees are depicted in Figure 4d-h. It is clearly that each concentration of aflatoxin B1 matches one phase singularity. The result corroborates the theory, the singularity on the curve is clear. This method mainly adds precision of measurement for phase singularity. Employing this method helps to avoid unstable measurements at the singularity, thereby improving measurement accuracy. Furthermore, compared to direct measurements, responses obtained from locations far away from the singularity exhibit greater stability and are less susceptible to interference. This leads to more accurate results and allows for solution derivation based on a set of measurement data, thereby reducing the impact of measurement errors on the final outcomes.

Table 2 shows the concentration of aflatoxin B1, corresponding incident polarization angle. Considering the define of sensitivity of traditional metasurface sensors [50–55], we are not able to take comparison of sensitivity with traditional methods directly. Compared to our previous metasurface sensor [56], the sensing indicator changes distinctly when the analyte is in trace levels. More importantly, compared to the change of amplitude and frequency, the incident polarization angle is an independent mechanical quantity which can be amplified by mechanical methods and geometrical optics. Currently, the measurement is also limited by the spectrometer, which is a challenging. However, we believe that with breakthroughs in technologies, the advantages of this method will surpass other sensors.

Table2 The concentration and polarization

| Concentration (μg/mL) | 0 | 0.001 | 0.01 | 0.1 | 1 |
|---|---|---|---|---|---|
| Polarization（degree） | 43.31 | 46.72 | 47.54 | 49.78 | 50.67 |

## Discussion

In summary, we have proposed a novel THz sensing method based on polarization-control phase singularity to address the current challenges faced by non-Hermitian metasurface sensors. We demonstrate the second-order non-Hermitian system using a metasurface and controllable linearly polarized THz wave. By introducing the concept of polarization-controlled losses, we establish a connection between two different types of physical quantities, the polarization of the incident light and the phase singularities of the coupled metasurface system. This method is not only well-suited in THz regime, but can also be extended to other frequencies including microwave (for more results please see supplementary information SN 7 and the attached video file), infrared, and visible. Furthermore, we develop a sensor characterization method based on the phase expression of the system's eigenvalues, using the polarization rotation angle as an indicator. We also propose a method to locate phase singularities, which circumvents challenges associated with directly measuring the system response, significantly enhancing the sensing performance. Note that without sacrificing generality, here we have opted for classical meta-structures in designing the sensor for the proof of concept, and the sensing performance can be further improved through the integration of advanced designs, nonlinear effects, and higher-order coupling. We aspire for this sensing method to augment the application potential of non-Hermitian THz metasurface sensors, and in particular, enhance the stability of non-Hermitian sensor measurements.

*B* **33**, 038701 (2024).

## Declarations

**Funding**: This work was supported by:

Self-determined research funds of CCNU from the colleges basic research and operation of MOE (CCNU23CG016)

Fundamental Research Funds for the Central Universities (CCNU24JCPT011)

National Natural Science Foundation of China (Grants No. 12364045, No. 12264028, and No. 12304420),

Natural Science Foundation of Jiangxi Province (Grants No. 20232BAB201040 and No. 20232BAB211025),

Young Elite Scientists Sponsorship Program by JXAST (Grant No. 2023QT11, No. 2025QT04).

**Author contributions:**

Conceptualization: Hai Lin, Xintong Shi

Resources: Yun Shen, Hai Lin, Rongxin Tang

Investigation: Xintong Shi, Yanjie Wu, Le Li, Chenhui Zhao, Shouxin Duan

Visualization: Xintong Shi, Junyi Zhang, Shuyuan Xiao

Supervision: Hai Lin, Yun Shen, Shuyuan Xiao

Writing—original draft: Xintong Shi

Writing—review & editing: Hai Lin, Tingting Liu, Shuyuan Xiao

**Competing interests:** All other authors declare they have no competing interests

**Data and materials availability:** All data are available in the main text or the supplementary materials.